\documentclass[aps,amssymb, showpacs]{revtex4}
\usepackage[english]{babel}
\usepackage{bm}
\usepackage{epsfig}
\newcommand{\beq}{\begin{equation}}
\newcommand{\eeq}{\end{equation}}
\newcommand{\ben}{\begin{eqnarray}}
\newcommand{\een}{\end{eqnarray}}

\newcommand{\noi}{\noindent}
\newcommand{\p}{\partial}
\newcommand{\h}{\hbar}

\begin{document}

\title{Quantum Mechanics of a Charged Particle in an Axial Magnetic Field}
\author{Ashok Das}
\affiliation{Department of Physics and Astronomy, University of
Rochester, Rochester, NY, USA}
\author{J. Frenkel and S. H. Pereira}
\affiliation{Instituto de F\'{i}sica, Universidade de S\~{a}o Paulo, S\~{a}o Paulo, Brazil}
\author{J. C. Taylor}
\affiliation{DAMTP, Centre for Mathematical Sciences, Cambridge University, Cambridge, UK}

\bigskip
\begin{abstract}
 We study some aspects of the quantum theory of a charged particle
moving in a time-independent, uni-directional magnetic field.
When the field is uniform, we make a few clarifying remarks on the use
of angular momentum eigenstates and momentum eigenstates with the
diamagnetism of a free electron gas as an example. When the field is
non-uniform but weakly varying, we discuss both perturbative and
non-perturbative methods for studying a quantum mechanical system. As an application, we derive the
quantized energy levels of a charged particle in a Helmholtz coil,
which go over to the usual Landau levels in the limit of a uniform field. 
\end{abstract}


\maketitle

\section{Introduction}

Although the quantum theory of a charged particle moving in a
uni-directional constant magnetic field has been studied over many decades (for some
text-book treatments, see for example \cite{magnetic}),
we would like to make some clarifying remarks which we believe are
useful. In studying this problem, one has to make a choice of the
gauge. Different gauge choices, of course, lead to a trivial phase
redefinition of the wave function. Moreover, corresponding to specific
gauges, one chooses either a wave function which is an eigenstate of
the momentum operator (which has continuous eigenvalues) or that of
the angular momentum operator (which has discrete eigenvalues). In
section II we point out that such a choice is unessential and derive
an explicit relation between the two bases. This is then used in the
treatement of diamagnetism of a free electron gas with angular
momentum eigenstates. In section III, we study the case when the magnetic field has a weak
$z$-dependence (where the $z$-axis is in the direction of the field).
Then, there must be also a weak radial field and we work under
conditions where it is consistent to treat this component of the field
perturbatively.
We find the energy eigenvalues for certain models of the $z$-dependence.
These models could approximate the field between two parallel
circular currents. In section IV, we discuss the form of the
wave-function for large quantum numbers associated with the
$z$-motion, using a non-perturbative
approach in a slowly varying field. As an application, we study in section V the
structure of the quantized energy levels of a charged particle in a
Helmholtz coil, which are specified by two quantum numbers. The first
is associated with the motion in the $(x,y)$ plane, while the second
characterizes the motion in the $z$-direction. This energy spectrum
may be regarded as a generalization of the Landau levels
\cite{magnetic} to the case of a slowly varying magnetic field. A
brief conclusion is presented in section VI.

\def\L{ \Lambda}
\def\p{\partial}
\def\f{\phi}
\def\m{\mu}
\def\n{\nu}
\def\l{\langle}
\def\r{\rho}
\def\D{{\cal D}}
\def\q{\psi}
\def\g{\gamma}
\def\B{{\cal B}}
\def\E{{\cal E}}

\section{Uniform magnetic field}

We take the field to be in the $z$-direction, and to have $z$-component $B$.
We take the particle's charge and mass to be $e$ and $m$.
Two gauges for the vector potential which are commonly used are
\begin{equation}
 {\bf A}= (-By/2,Bx/2,0) \label{a1}
\end{equation}
and
\begin{equation}
 {\bf A}= (-By, 0, 0).
\end{equation}
But the difference between these two gauges can only be trivial,
since the wave-functions just differ by the phase factor
\begin{equation}
\omega=\exp[i eBxy/(2c\hbar )].
\end{equation}
Therefore we might just as well choose (\ref{a1}).

For shortness, we define
\begin{equation}
 \B = eB/c
\end{equation}
The Hamiltonian in gauge (\ref{a1}) is
$$
H={1\over 2m}[(p_x-\B y/2)^2+(p_y+\B x/2)^2+p_z^2]
$$
\begin{equation}
={1\over 2m}[p_x^2+p_y^2+p_z^2+\B^2(x^2+y^2)/4 +\B L]=H'+p_z^2/2m,\label{a5}
\end{equation}
where
\begin{equation}
p_x=-i\hbar {\p \over \p x},~~p_y=-i\hbar {\p \over \p y},~~
p_z=-i\hbar {\p \over \p z},
\end{equation}

\begin{equation}
L=xp_y-yp_x\,.\end{equation}

We define also
\begin{equation} 
P_x=p_x+ \B y/2,~~ P_y=p_y- \B x/2. \label{a8}
\end{equation}
Then the operators $P_x,P_y$ and $L$ each commute with $H$ and $p_z$. But they don't
commute with each other:
\begin{equation}
[P_x,P_y]= i\hbar \B,~ ~[L,P_x]= i\hbar P_y,~ ~[L,P_y]=- i\hbar P_x
\end{equation}

We may choose the wave-function $\psi (x,y,z)$ to be an eigenfunction
of $p_z$ with eigenvalue $\hbar k_z$:
\begin{equation}
\psi(x,y,z)=e^{ik_z z}\phi(x,y).
\end{equation}
\begin{equation} 
H\psi=e^{ik_z z} [H'+\hbar^2k_z^2/(2m)]\phi,
\end{equation}
and from now on we will be concerned with $H'$ and $\phi(x,y)$.

We may choose $\phi$ to be an eigenfunction of any one of the three operators
$P_x,P_y,L$. We discuss the two cases (i) eigenfunction of $P_x$ (by
rotational invariance we can equally well choose it to be an
eigenstates of $P_y$) and (ii)
eigenfunction of $L$.
These wave-functions may be written in a more compact form if we define a magnetic length $l$ by
\begin{equation}
l= \sqrt{\hbar/\B}.
\end{equation}

In case (i), we have
\begin{equation}
\phi_{k_x,n}(x,y)= {1\over \sqrt{2\pi }}e^{ik_x x}e^{-i
xy/2l^2}\xi_n(y), \label{a13}
\end{equation}
where $\hbar k_x$ is the eigenvalue of $P_x$, and $\xi_n$ satisfies
\begin{equation}
{\hbar^2\over 2m}\left[ - {\p^2\over \p y^2}+ \{(y/l^2)-k_x\}^2 \right]\xi_n(y)=E'\xi_n(y),\label{a14}
\end{equation}
$E'$ being the eigenvalue of $H'$. This equation describes a simple
harmonic oscillator whose equilibrium point is shifted, and
the eigenfunctions have the form
\begin{equation}
 \xi_n(y)=l^{-1/2}u_n[(y-\eta)/l],
   \end{equation}
where $n$ is a positive integer,
\begin{equation}
\eta= 
l^2k_x\, ,~~E'={\hbar \B \over m}(n+{1\over 2} )\label{a16}
\end{equation}
and $u_n(w)$ is the real, normalized solution of
\begin{equation}
 \left[ -{\p ^2 \over \p w^2}+w^2 \right]u_n(w)=(2n+1)u_n(w).
\end{equation}

Next we take the case (ii), where we use energy eigenfunctions which are
also eigenfunctions of angular momentum
$L$ with eigenvalue  $\hbar M$. We denote these normalized
wave-functions by
\begin{equation}
\zeta_{n,M}(x,y).\label{a18}
\end{equation}

Eigenfunctions of type (i), in equation (\ref{a13}), are non-normalizable
and have a continuous degeneracy, whereas those of type (ii) in
(\ref{a18}) are normalized and have a discrete degeneracy.
Nevertheless, it should be possible to express each type in terms of the other.
To this end, we consider a superposition of the form
\begin{equation}
\chi_{n,f}=\int dk_x f(k_x)\phi_{k_x,n}(x,y)\label{a19}
\end{equation}
where $f$ is some function to be determined and 
\begin{equation}
\int dx dy \chi^*_{n',f}\chi_{n,f}=\delta_{nn'}\int dk_x |f(k_x)|^2 .
\end{equation}

In order to make (\ref{a19}) an eigenstate of $L$, we require
$$
\hbar M\chi_{n,f}=L\chi_{n,f}
$$
$$
=-i\hbar(2\pi l)^{-1/2}\int dk_x f(k_x) \left(x{\p \over \p y}-y{\p \over \p x}
\right)
$$
\begin{equation}
\times
\exp[ik_x x -i xy/2l^2]u_n[(y/l)-lk_x].\label{a25}
\end{equation}
We convert $x$ and $x^2$ into derivatives of $e^{ik_x x}$
and integrate by parts to put these derivatives onto $f$ and $u$.
In this way, we find that equation (\ref{a25}) is satisfied provided $f$
obeys
\begin{equation}
-{ 1\over l^2}{d^2 f\over d k_x^2}+l^2k_x^2 f  = [2(n-M)+1]f.
\end{equation}
This equation is consistent, and its normalised solution is just
\begin{equation}
f(k_x)=l^{1/2} u_{n-M}[lk_x ].
\end{equation}
Thus, finally, the required eigenfunction of $L$ is determined to be 
\begin{equation}
\zeta_{n,M}(x,y)
=l^{1/2}
\int dk_x u_{n-M}(lk_x)\phi_{k_x,n}(x,y).\label{a28}
\end{equation}

Using the ortho-normality of the $u_n$ coefficients in (\ref{a28}), we can invert this
equation to find the momentum eigenfunctions in terms of the angular
momentum eigenfunctions:
\begin{equation}
\phi_{k_x,n}(x,y)=l^{1/2}\sum_{M=-\infty}^{n}u_{n-M}(lk_x)\zeta_{n,M}(x,y).
\end{equation}

The text-book treatment (see for example \cite{diamagnetism}) of the diamagnetism of a free electron gas is usually
formulated in terms of the eigenfunctions of momentum (\ref{a13}). The sample
is considered to have a finite size with dimensions $S_x, S_y$
in the $x$- and $y$-directions, in which case states contribute only for
\begin{equation}
|\eta | <S_y.
\end{equation}
So the number of states with a given energy and value of $k_z$
is
\begin{equation}
{\B S_x S_y \over 2 \pi \hbar}={S_xS_y \over 2\pi l^2}\label{a42}
\end{equation}
It ought to be possible to carry out the analysis using the angular momentum
eigen-functions (\ref{a18}). In that case, it is natural to consider a
cylindrical sample, axis along the $z$ axis, with radius $R$.
The asymptotic form of the  wave function (\ref{a18}), for large $\rho$,  is proportional to
\begin{equation}
 e^{iM\phi} \rho^{2n-M} \exp(-\rho^2/4l^2),
\end{equation}
and this has its maximum at
\begin{equation}
\rho_{max}=\sqrt{2(2n-M)}\,\,l.
\end{equation}
Thus we expect states to contribute for
\begin{equation}
2n-M < {R^2 \over 2l^2}.
\end{equation}

Then, for given $n$, the  range of $M$ is
\begin{equation}
2n-{R^2\over 2l^2}\leq M \leq n,
\end{equation}
so that the number of possible values of $M$ is
\begin{equation}
 {R^2 \over 2l^2}-n. \label{a47}
\end{equation}
The relevant values of $n$ are of order $(mkTl^2/\hbar^2)$ which is small
compared to $(R^2 /l^2)$ for typical values of $B$, $R$ and $T$ (temperature). Thus (\ref{a47}) gives a factor
\begin{equation}
{\pi R^2\over 2 \pi l^2 }
\end{equation}
which is proportional to the area just as in (\ref{a42}).


\section{Weakly varying magnetic field}

We now allow the magnetic field to vary slowly with $z$, but we restrict
ourselves to motion near the $z$-axis. Along the axis, we use an expansion
\begin{equation}
B_z(z)=B_0[1+b_1 (z/a)^2+b_2 (z/a)^4+...]\label{a30}
\end{equation}
where $a$ is a characteristic length.

A realistic case would be a pair of two similar current loops, each perpendicular to
the common $z$-axis and of radius $a$, separated by a distance
$2d$. This system, known as the Helmholtz coil, is shown in Fig. 1.

\begin{figure}[htb]
\begin{center}
\epsfig{file=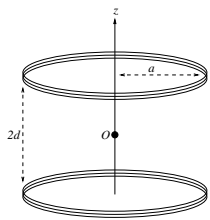, scale=0.5}\\
\end{center}
\caption{Schematic illustration of a Helmholtz coil.}
\end{figure}

Near the origin the field has an expansion
of the above form, where $B_0$ is determined by the current
carried by the loops and (in terms of $r=(d/a)^2$)
\begin{equation}
b_1= {3(4r-1) \over 2(r+1)^2},~~~b_2={15(1-12r+8r^2) \over 8(r+1)^4}.\label{a31}
\end{equation}

Off the axis, the magnetic field must have some radial component.
Up to second order (that is the $b_1$ term),
we take the vector potential to be
\begin{equation}
{\bf A}={1 \over 2}B_0[1+b_1(z^2-\rho^2/4)/a^2](-y,x,0). \label{a32}
\end{equation}
This form follows by rotational symmetry (if the field is generated by
currents in circles about the axis), together with the gauge
choice $\mbox{\boldmath$\nabla$}\cdot{\bf A}$$=0$ and Maxwell's equation
$\nabla^2 {\bf A}=0$. 

As the leading approximation, we keep just the $b_1 z^2$ term in (\ref{a30})
(we shall see that the $\rho^2$ term is effectively smaller).
We seek an energy eigenfunction of the Hamiltonian (\ref{a5}), with $\B$
now $z$-dependent, of the form
\begin{equation}
W(z)\Phi(x,y).
\end{equation}
Then the equation for $W$ is (we now define $l$ to be $\sqrt{\hbar/\B_0}$)
\begin{equation}
- {d^2W\over dz^2}+(2n+1){b_1z^2\over  a^2l^2}W={2m\Delta E\over  \hbar^2}W,
\end{equation}
where the second term on the left comes from $E'$ in (\ref{a14}) when $B$ is
$z$-dependent.
This is just the equation for a
 simple harmonic oscillator and, for $k_z=0$, the  energy associated with the $z$ motion is
\begin{equation}
\Delta E={\hbar^2 \over mal}\sqrt{ b_1 (2n+1)}(N+{1\over 2}).
\end{equation}

We can now treat the $z^4$ term in (\ref{a30}) and the $\rho^2$ term in (\ref{a32})
as perturbations. These terms are of the same order. This is because
the transverse size of the wave-function is of order $l$ whereas the
longitudinal size is of order $\sqrt{la}$. Note that, as an example, 
 for a magnetic flux density of 1 tesla, $l=2.5 \times 10^{-8} m$.

Treating the $z^4$ term by perturbation theory, we find an additional
energy
\begin{equation}
\Delta E' = {3\hbar^2 \over 4ma^2}{b_2\over b_1(2n+1)}(1+2N+2N^2).\label{a36}
\end{equation}
Note that this is independent
of $B$ and usually very small if $a$ is macroscopic. However, for large values of $N$, the contributions from the $b_2$ term and higher
 terms in (\ref{a30}) may be significant.

We also treat by perturbation theory the $\rho^2$ term in (\ref{a32}).
To this end
we require, inserting the $\rho^2$ term in (\ref{a32}) into the Hamiltonian (\ref{a5}),
\begin{equation}
\Delta E''=-{\hbar b_1\over 16ma^2 l^2}\int dx dy\phi^*_{n,M} (2L\rho^2+\hbar\rho^4/l^2)\phi_{n,M}.
\end{equation}
These expectation values can be worked out using:
\begin{equation}
\int dx dy \psi^*\rho^2 \psi= (2n-M+1)l^2,\label{a38}
\end{equation}
\begin{equation}
\int dx dy \psi^* \rho^4 \psi =[(2n-M+1)^2+2n(n-M)+2n-M+1]l^4.\label{a39}
\end{equation}
As a result, we have
\begin{equation}
\Delta E''=-{\hbar^2b_1\over 16 ma^2}[6n(n+1)-M(M+1)+2(1-nM)].\label{a40}
\end{equation}
This is independent of $B$ and, in general, is much smaller than
(\ref{a36}) for large values of $N$.


\section{Non-perturbative approach to a slowly varying field}

We have seen that when the quantum number associated with the
$z$-motion is large, $N >> n$, the dependence of $B_z$ upon $\rho$
may be neglected in first approximation. In this case, the vector
potential may be written as:
\begin{equation}
 {\bf A}= {1 \over 2} B_z(z)(-y,x,0) \,.\label{a49}
\eeq
\noi
This leads to the Schr\"odinger equation:
\beq
{\p^2 \psi \over \p \rho^2} + {1 \over \rho} {\p \psi \over \p \rho} -
\bigg({M^2 \over \rho^2}+ {e B_z M \over \hbar c} + {e^2 B_z^2
\over 4\hbar^2 c^2}\rho^2 \bigg)\psi + {\p^2 \psi \over \p z^2} = -
{2m \over \hbar^2} E \psi \,, \label{a50}
\eeq
\noi
where we have taken the energy eigenfunction to be also an
eigenfunction of the angular momentum $L_z$ with eigenvalue $\hbar
M$. Let us consider a solution of the form:
\beq
\psi(\phi, \rho, z) = F(\phi, \rho, z)f(z) \label{a51}
\eeq
\noi
and assume, since the field is slowly varying, that:
\beq
\bigg| {\p F \over \p z} \bigg| <<  \bigg| {\p F \over \p \rho} \bigg|
\,. \label{a52}
\eeq
\noi
Then, the Schr\"odinger equation (\ref{a50}) takes the form:
\beq
{1 \over F} \bigg[ {\p^2 F \over \p \rho^2} + {1 \over \rho}{\p F
\over \p \rho}\bigg] - {M^2 \over \rho^2} - {eB_zM \over \hbar c}-
{e^2 B_z^2 \over 4 \h^2 c^2}\rho^2 = - \bigg[ {1 \over f}{d^2f \over d z^2}
+ {2m \over \h^2} E \bigg]\,. \label{a53}
\eeq
\noi
The right hand side of (\ref{a53}) depends on $z$ only, so that we
have:
\beq
{d^2f \over d z^2} + {2m \over \h^2} E f = {2m \over \h^2} E'(z) f
\label{a54}
\eeq
\noi
and
\beq
{\p^2 F \over \p \rho^2} + {1 \over \rho}{\p F \over \p \rho} - \bigg(
{M^2 \over \rho^2} + {eB_z M \over \h c} + {e^2 B_z^2 \rho^2 \over 4
\h^2 c^2}\bigg)F = -{2m \over \h^2} E'(z) F \,. \label{a55}
\eeq
\noi
But we know how to solve (\ref{a55}), in which $z$ is just a
parameter. Comparing with (\ref{a16}), we see that:
\beq
E'(z) = {e \h \over mc}(n+ {1 \over 2}) B_z(z) \,. \label{a56}
\eeq
\noi
Thus, (\ref{a54}) may be written as:
\beq
{d^2 f \over d z^2}+{2m\over \h^2} \bigg[ E - {e \h (2n+1)\over 2mc}
B_z(z)\bigg] f = 0 \,, \label{a57}
\eeq
\noi
which is just an ordinary differential equation. 

We can now check the
assumption (\ref{a52}). The solutions of (\ref{a55}) have the form
(\ref{a18}), namely:
\beq
F(\phi, \rho, z) = \zeta_{n,M}(\phi, \sqrt{{eB_z\over 2\h c}}\,\rho)
\,. \label{a58}
\eeq
\noi
Hence, the condition (\ref{a52}) requires that:
\beq
{\rho\over  B_z}{dB_z\over dz} << 1 \,. \label{a59}
\eeq
\noi
Although this cannot be satisfied for all $\rho$, we note that the
solution (\ref{a58}) contains the factor:
\beq
\exp[-{eB_z\over 4\h c}\rho^2]\,,\label{a60}
\eeq
\noi
so that typical values of $\rho$ are of order $\sqrt{\h c/eB_z}$. Thus, we expect (\ref{a59}) to be valid for:
\beq
\sqrt{{\h c \over e}}{1\over B_z^{3/2}}{dB_z\over dz}<< 1\,. \label{a61}
\eeq
\noi
This can be satisfied provided we take $B_z$ to be varying slowly
enough. For instance, if $B_z(z)$ has the form (\ref{a30}) with $|z|$ of order $a$, then the left hand side of (\ref{a61}) is of order $l/a$, which is very small for typical values of $B_0$ and $a$.


\section{Quantized energy levels in a Helmholtz coil}

A simple example of a magnetic mirror is provided by the pair of
current loops
shown in Fig. 1. Then, as can be seen from equations (\ref{a30}) and
(\ref{a31}) for $2d > a$, the magnetic field can increase enough in
the region near the current loops, so that a charged particle may
eventually be reflected out of this region towards the center of the
coil. The classical restoring force which may confine the particle along the
$z$-axis is provided by the $\rho$-component of the magnetic field:
\beq
{\bf B}(z) = -{1\over 2}\rho {\p B_z\over \p z}\hat{\mbox{\boldmath$\rho$}} + B_z
\hat{{\bf z}} \,. \label{a62}
\eeq 
which can be derived from (\ref{a49}) using ${\bf B} =
\mbox{\boldmath$\nabla$}\times{\bf A}$.

In this section, we will discuss the quantized energy levels of a
charged particle in a Helmholtz coil. To this end, we will analyse the basic equation
(\ref{a57}), where
\beq
B_z(z) = {1 \over 2} B_0 (d^2 + a^2)^{3/2}\bigg\{ {1 \over[(z+d)^2 +
a^2]^{3/2}} + {1 \over[(z-d)^2 +
a^2]^{3/2}} \bigg\} \,. \label{a63}
\eeq
The general behavior of this field, for $2d>a$, is shown in Fig. 2. 
\begin{figure}[htb]
\begin{center}
\epsfig{file=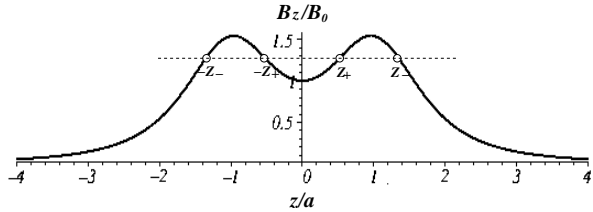, scale=0.3}\\
\end{center}
\caption{Pattern of the field $B_z(z)$ in a Helmholtz coil. Circles
indicate the classical turning points.}
\end{figure}

Since (\ref{a57}) was derived under the assumption that $N>>1$, our
analysis will be more appropriate in the semiclassical regime. This may
be studied conveniently in the WKB approximation, which is
particularly useful since we are dealing with a slowly varying
potential:
\beq
V(z) = { e\h \over 2mc}(2n+1)B_z(z)\,.\label{a64}
\eeq
One of the most interesting results arising from the WKB method is a
semiclassical estimate for the quantized energy levels in a
potential. Matching the WKB wave function at each of the classical
turning points, which are determined by the
relation $E=V(z)$, leads to the Bohr-Sommerfeld quantization
condition:
\beq
\int_{-Z_+}^{Z_+}\sqrt{2m[E - V(z)]}dz = (N+{1\over 2})\pi \h\,,
\label{a65}
\eeq
where the classical turning points for bound motion, $\pm Z_+$, are
situated inside the well as shown in Fig. 2. (We neglect, in first approximation, the very small
probability of tunneling through the potential barrier.)

In general, equation (\ref{a65}) is rather complicated and can
only be solved numerically. However, when $d/a$ is somewhat larger
than $1/2$, it may be solved in closed form, since in this case $z/a$
will be effectively of order $1/2$ or smaller. Then, we can expand the
potential $V(z)$ [see (\ref{a63}) and (\ref{a64})] up to terms which
are of quartic order in $z/a$, as shown in equations (\ref{a30}) and
(\ref{a31}). In this approximation, the quantization condition
(\ref{a65}) may be written in the form:
\beq
\int_{-Z_+}^{Z_+} \sqrt{\E -1 - b_1 \big({z\over a}\big)^2 -
b_2\big({z\over a}\big)^4}dz = \pi \sqrt{{c\h \over (2n+1)e B_0}}
(N+{1\over 2})\,, \label{a66}
\eeq
where the dimensionless parameter $\E$ is defined by:
\beq
\E = {2mc\over (2n+1)e\h B_0}E\,.\label{a67}
\eeq
We can now determine explicity the positions of the
classical turning points, which are given by:
\beq
Z_\pm = a\bigg[ {-b_1 \pm \sqrt{b_1^2 + 4b_2 (\E -1)}\over
2b_2}\bigg]^{1/2}\,, \label{a68}
\eeq
where $\pm Z_-$ are the turning points for unbound motion, which are
situated outside the well as shown in Fig. 2. Then, the integral appearing in (\ref{a66}) can be
evaluated in closed form in terms of the complete elliptic integrals
of the first (${\bf F}$) and second (${\bf E}$) kind \cite{tables}, with the result:
\beq
{2\sqrt{-b_2}\over 3a^2} Z_- \bigg[ \big(Z_+^2 - Z_-^2\big) {\bf
F}\bigg({Z_+\over Z_-}\bigg) +\big(Z_+^2 + Z_-^2\big) {\bf
E}\bigg({Z_+\over Z_-}\bigg)\bigg]=\pi \sqrt{{c\h\over
(2n+1)eB_0}}(N+{1\over 2})\,. \label{a69}
\eeq
This is a transcendental equation which determines implicitly the
energy in terms of the quantum number $N$. The energies of the bound states have an upper bound $E^{max}$ which corresponds
to the maximum value of the potential. At this energy, $Z_+ = Z_-$, so
that using (\ref{a67}) and (\ref{a68}) we obtain:
\beq
E^{max}_n= {e\h B_0\over 2mc}(2n+1)\bigg(1-{b_1^2\over
4b_2}\bigg)\,. \label{a70}
\eeq

When $Z_+ = Z_-$, the equation
(\ref{a69}) simplifies considerably and fixes the
maximum value of the quantum number $N$, which is given by the relation:
\beq
N_{max}+ {1\over 2}= {\sqrt{2(2n+1)}\over 3\pi}{b_1^{3/2}\over |b_2|}{a\over
l}\,, \label{a71}
\eeq
where $l=\sqrt{c\h / e B_0}$ is the magnetic length. Hence, $N_{max}$ is
in general a very large number, which is in acordance with our
previous assumption.

A more explicit relation giving $E$ as a function of $N$ can be obtained
by solving numerically the equation (\ref{a69}). As an example, the numerical
solution for $n=0$ and $d/a=3/5$ is shown in Fig. 3. Here, the
numerical values of $\E^{max}=1.028$ and $N_{max}=1.12\times 10^6$
are in good agreement with the corresponding results obtained from the
closed form expressions (\ref{a70}) and (\ref{a71}).  One can see that the numerical solution
can be fitted reasonably well, for large quantum numbers, by
a simple phenomenological form like $N^{5/6}$.

\begin{figure}[htb]
\begin{center}
\epsfig{file=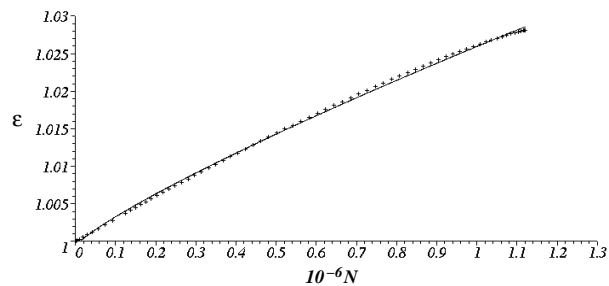, scale=0.3}\\
\end{center}
\caption{Numerical solution (crossed) of Eq. (\ref{a69}) for the
energy spectrum and plot (solid) of the form $N^{5/6}$ for large $N$.}
\end{figure}
Finally, we note from (\ref{a30}) and (\ref{a31}) that for $d/a=1/2$, $b_1=0$ and $b_2=-144/125$. Then, the
system of the two current loops provides a practically uniform field in the
central region of the Helmholtz coil. In this case, (\ref{a70})
reduces to the well known expression for the Landau levels which occur
in a constant field. Therefore, one may regard the
quantized energies described by (\ref{a69}) and (\ref{a70}),
as an extension of the Landau levels to the case of a slowly varying magnetic field.

\section{Conclusion}

In this paper, we have discussed several aspects concerning the
quantum behavior of a charged particle in a static magnetic field. We
have treated the issue of the relation between the choice of gauge
and the choice of the diagonal operators which commute with the
Hamiltonian of the system. We have also developed some approaches
which may be useful for physical applications in a slowly varying
magnetic field. These methods have been applied to study the quantized
energy levels of a charged particle in a weakly varying field. Such a
field may be present, for example, in a magnetic mirror like the
Helmholtz coil. We have shown that this energy spectrum represents an
interesting extension of the well known Landau levels which occur in an
uniform magnetic field.

\vspace{0.5cm}

This work was supported in part by US DOE Grant number DE-FG
02-91ER40685, by CAPES, CNPq and FAPESP, Brazil.

\end{document}